\documentclass[12pt,final]{l4dc2020} 
\usepackage{comment}
\usepackage{times}
\usepackage{sidecap}
\usepackage{ifthen}
\usepackage{graphicx}
\usepackage{subcaption}
\DeclareMathOperator{\Hcal}{\mathcal{H}}
\usepackage{wrapfig}

\title[Improving I-O Linearizing Controllers for Bipedal Robots via RL]{Improving Input-Output Linearizing Controllers for Bipedal Robots via Reinforcement Learning}

\author{\Name{Fernando Casta\~neda}$^1$ \Email{fcastaneda@berkeley.edu}\\
  \Name{Mathias Wulfman}$^1$ \Email{mathias$\_$wulfman@berkeley.edu}\\
  \Name{Ayush Agrawal}$^1$ \Email{ayush.agrawal@berkeley.edu}\\
  \Name{Tyler Westenbroek}$^2$
  \Email{westenbroekt@berkeley.edu}\\
  \Name{Claire J. Tomlin}$^2$
  \Email{tomlin@berkeley.edu}\\
  \Name{S. Shankar Sastry}$^2$
  \Email{shankar$\_$sastry@berkeley.edu}\\
  \Name{Koushil Sreenath}$^1$ \Email{koushils@berkeley.edu}\\
  \addr $^1$ Department of Mechanical Engineering, University of California at Berkeley, USA\\
  \addr $^2$ Department of Electrical Engineering and Computer Sciences, University of California at Berkeley, USA
  }
  
\begin{document}

\maketitle

\begin{abstract}%
The main drawbacks of input-output linearizing controllers are the need for precise dynamics models and not being able to account for input constraints. Model uncertainty is common in almost every robotic application and input saturation is present in every real world system. 
In this paper, we address both challenges for the specific case of bipedal robot control by the use of reinforcement learning techniques. Taking the structure of a standard input-output linearizing controller, we use an additive learned term that compensates for model uncertainty. Moreover, by adding constraints to the learning problem we manage to boost the performance of the final controller when input limits are present.
We demonstrate the effectiveness of the designed framework for different levels of uncertainty on the five-link planar walking robot RABBIT.
\end{abstract}


\section{Introduction}
\label{sec:Introduction}
\subsection{Motivation}
\hspace{0.5cm} Research on humanoid walking robots is gaining in popularity due to the robots' medical applications as  exoskeletons for people with physical disabilities and their usage in dangerous disaster and rescue missions. Model-based controllers have traditionally been applied to obtain stable walking controllers but, in general, they heavily rely on having perfect model knowledge and unlimited torque capacity. In this paper we take a data-driven approach to address these two topics of current research interest which still constitute challenges in bipedal robot control: uncertainty in the dynamics and input saturation.

\subsection{Related work}
\hspace{0.5cm} Input-output linearization is a nonlinear control technique that can be used to get the outputs of a nonlinear system to track desired reference trajectories in a simple manner. By introducing an appropriate state transformation, this control technique permits rendering the input-output dynamics linear. Afterward, linear systems control theory can be used to track the desired outputs. However, input-output linearization requires precise knowledge of the system's dynamics, which directly conflicts with the fact that actual systems' dynamics might have nonlinearities that can be extremely challenging to model precisely. Several efforts have been made to address this issue, using different methods including robust and adaptive control techniques \citep{article, Sastry:1989:ACS:63437, number9, number10} or, more recently, data-driven learning methods \citep{taylor2019episodic, westenbroek2019feedback}. This paper will take the later approach to address this challenge, specifically combining reinforcement learning (RL) and the Hybrid Zero Dynamics (HZD) method for getting bipedal robots to walk.
\smallbreak

The high nonlinearity, underactuation and hybrid nature of bipedal robotic systems pose additional problems that need to be addressed. The virtual constraints and HZD methods \citep{Grizzle2001, WESTERVELT2002, Westervelt2003, MorisGrizzle} provide a systematic approach to designing asymptotically stable walking controllers if there is full model knowledge. These methods have been very successful in dealing with the challenging dynamics of legged robots, being able to achieve fast enough convergence to guarantee stability over several walking steps. By the HZD method, a set of output functions is chosen such that, when they are driven to zero, a time-invariant lower-dimensional zero dynamics manifold is created. Stable periodic orbits designed on this lower-dimensional manifold are also stable orbits for the full system under application of, for instance, input-output linearizing \citep{Sreenath}, or control Lyapunov function (CLF) based controllers \citep{GrizzleSreenath}. The later is based on solving online quadratic programs, whereas the former approach does not rely on running any kind of online optimization. The CLF-based method has also been successful in taking into account torque saturation \citep{torque}, but it assumes perfect model knowledge too. In fact, taking input saturation into account is of major importance and not doing it is one of the main disadvantages of input-output linearization controllers that is often overlooked.
\smallbreak
In this work, we build on the formulation proposed in \cite{westenbroek2019feedback} wherein policy optimization algorithms from the RL literature are used to overcome large amounts of model uncertainty and learn linearizing controllers for uncertain robotic systems. Specifically, we extend the framework introduced in \cite{westenbroek2019feedback} to the class of hybrid dynamical systems typically used to model bipedal robots using the HZD framework. Unlike the systems considered in \cite{westenbroek2019feedback}, here we must explicitly account for the effects of underactuation when designing the desired output trajectories for the system to ensure that it remains stable. Additionally, we demonstrate that a stable walking controller can be learned even when input constraints are added to the system.
By focusing on learning a stabilizing controller for a single task (walking), we are able to train our controller using significantly less data than was used in \cite{westenbroek2019feedback}, where it was trained to track all possible desired output signals.

\subsection{Contributions}
The contributions of our work thus are:
    \vspace{-1pt}
\begin{itemize}
    \item We extend the work in \cite{westenbroek2019feedback} to the case of hybrid, underactuated bipedal robots with input constraints.
    \vspace{-2pt}
    \item We directly address the challenge of dealing with a statically unstable underactuated system, designing a new training strategy that uses a finite-time convergence feedback controller to track desired walking trajectories.
    \vspace{-2pt}
    \item We perform Poincar\'e analysis to claim local exponential stability of our proposed RL-enhanced input-output linearization controller in the presence of torque saturation.
\end{itemize}
\subsection{Organization}
\hspace{0.5cm} The rest of the paper is organized as follows. Section \ref{Sec2} briefly revisits hybrid systems theory for walking and input-output linearization. Section \ref{Sec3} develops the proposed RL framework that improves the input-output linearizing controller when there is a mismatch between the model and the plant dynamics. Section \ref{Sec4} presents simulations on perturbed models of RABBIT, a five-link planar bipedal robot. Finally, 
Section \ref{Sec5} provides concluding remarks.

\section{Input-Output Linearization of Bipedal Robots}
\label{Sec2}
\subsection{Model Description}
\hspace{0.5cm} Bipedal walking is represented as a hybrid model with single-support continuous-time  dynamics  and double-support discrete-time impact dynamics  \eqref{eq:cdt}, with $x \in \mathbb{R}^{2n}$ being the robot  state, and
$u \in \mathbb{R}^m$ the control inputs. $x^-$ and $x^+$ represent the state before and after impact, respectively, with $\mathcal{S}$ being the switching surface when the swing leg contacts the ground and $\Delta$ being the discrete-time impact map. The constrained continuous-time dynamics are represented in the manipulator form \eqref{eq:ct}, where $q \in \mathbb{R}^{n}$ is the vector containing the generalized system's coordinates, $D(q)$ is the inertia matrix of the system, $C(q,\dot{q})$ is the matrix representing the centripetal and Coriolis effects, $G(q)$ is the gravitation terms vector, $B(q)$ is the motor torque matrix, $J(q)$ is the Jacobian of the stance foot and $\lambda$ is the ground contact forces vector. The state variables are $x=[q,\dot{q}]^{\top}$.

\begin{center}
\hspace{-15pt}
\begin{tabular}{  l | r }
\begin{tabular}{p{6cm}}
  \vspace{-25pt}
\begin{equation}
 \Hcal=\left\lbrace
      \begin{aligned}
        &\dot{x} = f(x)+g(x)u, &      &  x^{-} \notin \mathcal{S},\\
        &x^{+} = \Delta ( x^{-}),&    &   x^{-} \in \mathcal{S}.\\
      \end{aligned}
    \right.
    \label{eq:cdt}
    \vspace{-25pt}
\end{equation} \end{tabular} & \begin{tabular}{p{8cm}}
  \vspace{-25pt}
  \begin{equation}
  \left\{
  \begin{array}{l}
   D(q)\ddot{q}+C(q,\dot{q})\dot{q}+G(q) = B(q)u+J^{{\top}}(q)\lambda, \\
   J(q)\ddot{q}+\dot{J}(q,\dot{q})\dot{q} = 0. \\
  \end{array}
  \right.
  \label{eq:ct}
  \vspace{-25pt}
\end{equation} 
\end{tabular}
\end{tabular}

\end{center}

\subsection{Input-Output Linearization}
\label{sec:IO}
\hspace{0.5cm} The output function $ y : \mathbb{R}^{2n} \rightarrow \mathbb{R}^m $ is defined to represent the walking gait. 
Supposing $y$ has a vector relative degree two ---meaning that the first derivative of $y$ does not depend on the inputs but the second derivative does--- the second derivative of $y$ can be written as:
\begin{equation}
    \ddot{y}  = L_f^2y(x)+ L_gL_fy(x)u. \label{eq:sys_lie}
\end{equation} \smallbreak
The functions $L_f^2y$ and $L_gL_fy$ are known as second order Lie derivatives. More information about Lie derivatives and how to compute them can be found in \cite{lie}. 
Moreover, using the method of Hybrid Zero Dynamics (HZD) the output function and its first derivative are driven to zero, imposing “virtual constraints” such that the system evolves on the lower-dimensional zero dynamics manifold, given by $ Z=\{ x \in \mathbb{R}^{2n} |\ y(x)=0, \ \dot{y}(x)=0\}.$ If the vector relative degree is well-defined, then $L_gL_fy(x) \not= 0$ $\forall$ $x\in D$, with $D\subset \mathbb{R}^{2n}$ being a compact subset of the state space containing the origin. Since $L_gL_fy$ is nonsingular in $D$, we can use the input-output linearizing control law:
\begin{equation}
    u(x)=L_gL_fy^{-1}(x)(-L_f^2y(x) +v),
\end{equation}
which yields $ \ddot{y}=v$, where $v$ is a virtual input.

Suppose a state transform $\Phi : x \rightarrow (\xi, z)$, with $\xi =[y,\dot{y}]^{\top} $ and $z \in Z$. Then, the closed-loop dynamics become a linear time-invariant system on $\xi$ and the zero-dynamics on $z$:
\vspace{-10pt}
 \begin{center}
 \begin{equation}
\begin{tabular}{  l c }
\begin{tabular}{p{5cm}}
\vspace{-10pt}

  \begin{equation*}
 \left\lbrace
      \begin{aligned}
         \dot{\xi} & = A\xi +Bv,\\
         \dot{z} & = p(\xi,z),
      \end{aligned}
    \right.
    \vspace{-10pt}
\end{equation*} \end{tabular} & 
\hspace{-1cm}with 
$A = \begin{bmatrix}
0_{m \times m} & I_{m} \\
0_{m \times m} & 0_{m \times m}
\end{bmatrix}$ and $B = \begin{bmatrix}
0_{m \times m} \\
I_{m} 
\end{bmatrix}$.
\\

\end{tabular}
\end{equation}
\end{center}
We define $v$ following \cite{westervelt2018feedback}:
\vspace{-15pt}
 \begin{center}
\begin{tabular}{  l c }

$v(\xi)=\frac{1}{\epsilon^2}\psi_a(y,\epsilon \dot{y})$, & with 
\begin{tabular}{p{11cm}} 
\vspace{-15pt}
\begin{equation}
 \left\lbrace
      \begin{aligned}
         \psi_a(y,\epsilon \dot{y}) &= - sign(\epsilon \dot{y})|\epsilon \dot{y}|^a - sign(\phi_a(y,\epsilon \dot{y}))|\phi_a(y,\epsilon \dot{y})|^{\frac{a}{2-a}},\\
         \phi_a(y,\epsilon \dot{y})&=  y  +\frac{1}{2-a}sign(\epsilon \dot{y})|\epsilon \dot{y}|^{2-a} ,
      \end{aligned}
    \right.
    \label{v_eq}
    \vspace{-25pt}
\end{equation} 
\end{tabular}
\end{tabular}
\end{center}
such that $v$ ensures finite time convergence to $Z$ and $\epsilon$ controls the rate of convergence.

\section{Reinforcement Learning for Uncertain Dynamics}
\label{Sec3}

\hspace{0.5cm} In this section, we study the case in which there is a mismatch between the model and the actual plant dynamics. Now, plant and model are represented by:

\begin{center}

\begin{tabular}{  c | c }
  (Unknown) Plant Dynamics & (Known) Model Dynamics \\ \hline
\begin{tabular}{p{5cm}}
  \vspace{-15pt}
\begin{equation}
  \left\lbrace
      \begin{aligned}
         \dot{x}  & = f_p(x)+g_p(x)u,\\
         y & = h_p(x),
      \end{aligned}
    \right.
    \vspace{-15pt}
\end{equation} \end{tabular} & \begin{tabular}{p{5cm}}   \vspace{-15pt}
 \begin{equation} 
 \left\lbrace
      \begin{aligned}
         \dot{x}  & = f_m(x)+g_m(x)u,\\
         y & = h_m(x).
      \end{aligned}
    \right.
    \vspace{-15pt}
\end{equation} \end{tabular}
\end{tabular}

\end{center}

For our application we will be using the same output functions for plant and model, so we could actually set $h_p \equiv h_m$. Furthermore, we assume that both systems have vector relative degree two.
Defining an input-output linearizing controller on the model dynamics using the state dependent finite-time convergence feedback controller presented in \eqref{v_eq} for the additional input $v$ we get:
\begin{equation}
\label{eq:cont}
    u(x)=\left(L_{g_{m}}L_{f_m}h_m(x) \right)^{-1}(-L_{f_m}^2h_m(x) +v(x)).
\end{equation}
\smallbreak
However, if the mismatch between the model and the real dynamics is big enough, this controller may not manage to stabilize the plant. In order to address this issue we use an alternative control input:
\vspace{-15pt}
\begin{equation}
\label{eq:contRL}
    u_\theta (x)=\Big(L_{g_{m}}L_{f_m}h_m(x) \Big)^{-1}\Big(-L_{f_m}^2h_m(x)+v(x) \Big) +\alpha_\theta(x) v(x) + \beta_\theta(x),
\end{equation}
where  $\theta \in \mathbb{R}^k$ is a vector of parameters of a neural network that are to be learned. For a specific $\theta$, the policies $\alpha_{\theta}:\mathbb{R}^{n}\rightarrow\mathbb{R}^{m\times m},\ \beta_{\theta}:\mathbb{R}^{n}\rightarrow\mathbb{R}^{m}$ take the current state as input and serve to define an additive learned term that is affine in $v$. Note that $u_\theta$ maintains the structure of an input-output linearizing controller. Applying the new control law $u_\theta$, the second derivative of the plant's outputs can be rewritten as:
\vspace{-10pt}
\begin{equation}
\label{eq:ddy}
    \ddot{y}  = L_{f_p}^2h_p(x)+ L_{g_p}L_{f_p}h_p(x)\Bigg(\Big(L_{g_{m}}L_{f_m}h_m(x) \Big)^{-1}\Big(-L_{f_m}^2h_m(x)+v(x) \Big) +\alpha_\theta(x) v(x) + \beta_\theta(x)\Bigg).
\end{equation}
In \cite{westenbroek2019feedback}, $W_\theta$ is defined as the right hand side of the above equation, such that $\forall x \in \mathbb{R}^{2n},$ $\ddot{y}=W_\theta(x)$. The point-wise loss is then defined on $\mathbb{R}^{2n} \times \mathbb{R}^{k}$ as:
\vspace{-2pt}
\begin{equation}
\label{eq:loss}
    l(x,\theta) = ||v(x)-W_\theta(x) ||_2^2,
\vspace{-2pt}
\end{equation}
which provides a measure of how well the controller $u_\theta$ linearizes the plant at the state $ x $. Since the term $W_\theta$ present in the loss function depends on the unknown plant dynamics, we use a finite difference approximation of it by replacing this by the second derivative of the outputs of the plant.\smallbreak

Now, we will formulate our problem as a canonical RL problem \citep{SuttonBarto}. Even though only $\alpha_\theta$ and $\beta_\theta$ are learned, for the sake of simplicity let $\pi_\theta: x \mapsto \pi_\theta (x)$ be our policy taking the current state $x$ and returning the control action $u_\theta=\pi_\theta (x)$, and let the reward for a given state $x$ be $R(x,u_\theta)=-l(x,\theta) + R_e(x)$, where $R_e(x)$ is a penalty value if the state $x$ is associated with a fallen robot configuration or a bonus value otherwise. Then, we can define the learning problem
\begin{equation} 
\begin{aligned}
\label{eq:RLoptimization}
& \underset{\theta}{\text{max}}
& & \mathbb{E}_{x_0 \sim X_0, w \sim \mathcal{N}(0,\sigma^2)}   \int_0^T R(x(\tau),u_\theta (\tau))d\tau, \\
& \text{s.t.} & &  \dot{x}=f(x)+g(x)(\pi_\theta (x)+ w_t ),\\
& \ & &  u_{min}\leq \pi_\theta (x) \leq u_{max},
\end{aligned}
\end{equation}
where $X_0$ is the initial state distribution, $T >0 $ is the duration of the episode, $w$ is an additive zero-mean noise term and $u_{min}$ and $u_{max}$ are the torque limits.
An episode ends when the robot completes an entire step or when it falls. A discrete-time approximation of this problem can be solved using standard on-policy and off-policy RL algorithms. Note that our proposed controller \eqref{eq:contRL} with the chosen loss \eqref{eq:loss} and the inclusion of input constraints in the optimization \eqref{eq:RLoptimization} addresses the classical challenges of input-output linearization: model uncertainty and input constraints. From now on, we will call \textit{original IO controller} the one of \eqref{eq:cont} and \textit{RL-enhanced IO controller} the one of \eqref{eq:contRL}, with $\theta$ chosen by solving \eqref{eq:RLoptimization}.

\section{Simulation}
\label{Sec4}
\subsection{System Description}
\begin{SCfigure}[1.1]%
\centering
\subfigure[RABBIT]{%
\includegraphics[scale=0.21]{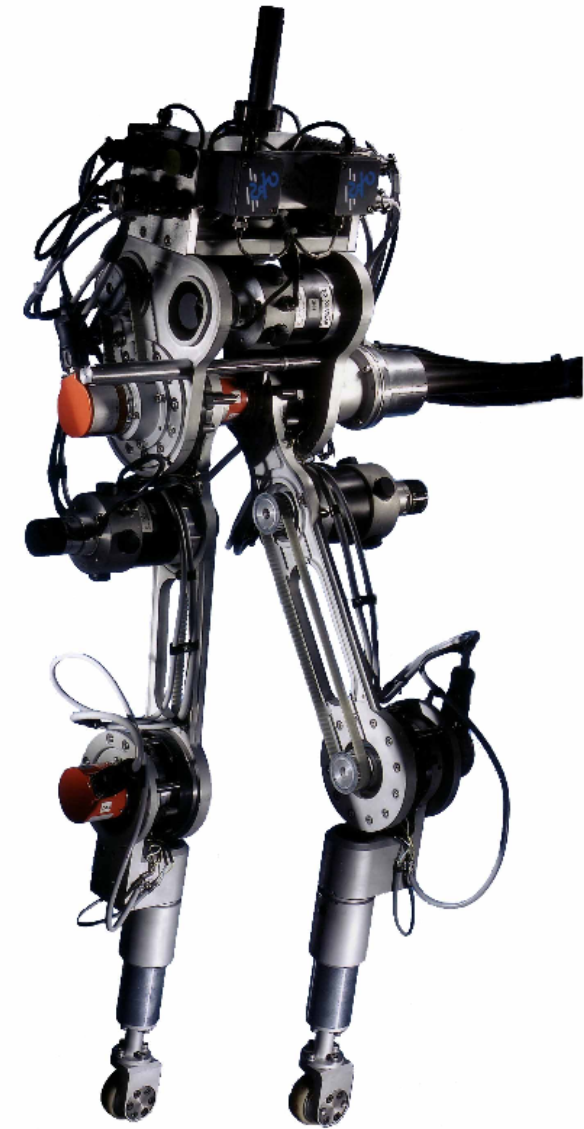}%
}
\qquad
\subfigure[Coordinate system]{%
\includegraphics[scale=0.21]{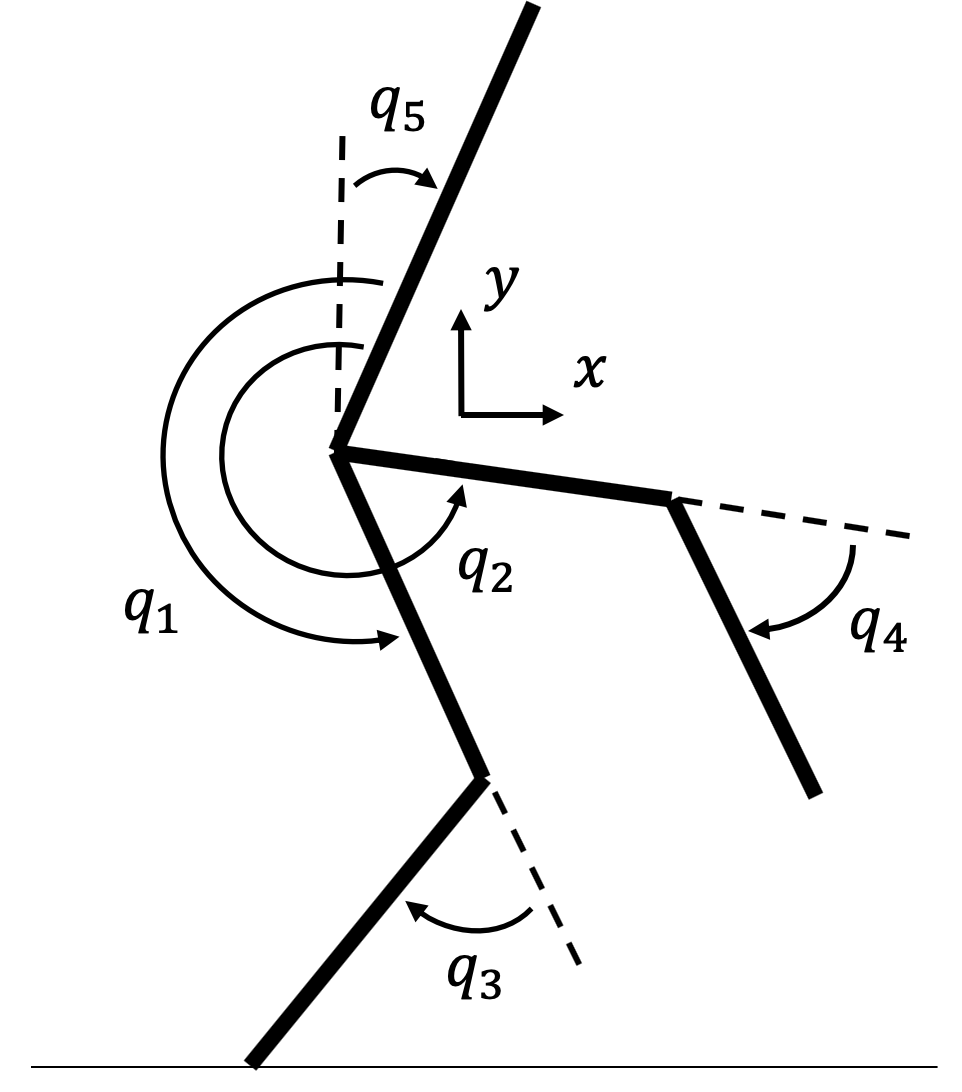}%
}
\caption{(a) RABBIT, a planar five-link bipedal robot with nonlinear, hybrid and underactuated dynamics. (b) $q_1$, $q_2$ are the relative stance and swing leg femur angles referenced to the torso, $q_3$, $q_4$ are the relative stance and swing leg knee angles, $q_5$ is the absolute torso angle in the world frame, and $x$ and $y$ are the position of the hip in the world frame. Here $q=[x, y, q_1, q_2, q_3, q_4, q_5]^{\top}$.}
\label{fig:Rabbit}
\vspace{-12pt}
\end{SCfigure}

\hspace{0.5cm} In order to numerically validate our method, we use a model of the five-link planar robot RABBIT \citep{Chevallereau03rabbit:a}, wherein the stance phase is parametrized by a suitable set of coordinates (Figure \ref{fig:Rabbit}). RABBIT is a 7 Degrees-of-Freedom (DOF) underactuated system with 4 actuated DOF, with the actuators being located at the four joints (the two hip joints and the two knee joints). The dynamics of this 14-dimensional system are extremely coupled and nonlinear.
\makeatletter
 \let\Ginclude@graphics\@org@Ginclude@graphics 
\makeatother
\vspace{-5pt}
\subsection{Reference Trajectory Generation}
\hspace{0.5cm} In order to generate a reference trajectory offline, we use the Fast Robot Optimization and Simulation Toolkit (FROST) \citep{FROST}. The four actuated DOF ($q_1$, $q_2$, $q_3$ and $q_4$) are virtually constrained to be B\'ezier Polynomials of the stance leg angle $\theta = q_5 + q_1 + \frac{q_3}{2}$, which is monotonically increasing during a walking step. This way, the trajectory that has been generated is time-invariant, which makes the controlled system more robust to uncertainties \citep{westervelt2018feedback}. Taking the difference between the actual four actuated joint angles and the desired ones (coming from the reference trajectory) as output functions $y$, the system is input-output linearizable with vector relative degree two. Consequently, we can use the \textit{RL-enhanced IO controller} $u_\theta$ presented in the previous section.
\smallbreak
We train our controller using a Deep Deterministic Policy Gradient Algorithm (DDPG) \citep{ddpg}. DDPG is used to tune the parameters of the actor and critic feedforward neural networks. They each have two hidden layers of widths 400 and 300 and ReLU activation functions.
The actor neural network maps 14 observations, which are the states of the robot, to 20 outputs corresponding to the $4 \times 4$ $\alpha_\theta$ and the $4 \times 1$ $\beta_\theta$.

\subsection{Model-Plant Mismatch and Torque Saturation Results}
\begin{figure}
    \centering
    \includegraphics[scale=0.91]{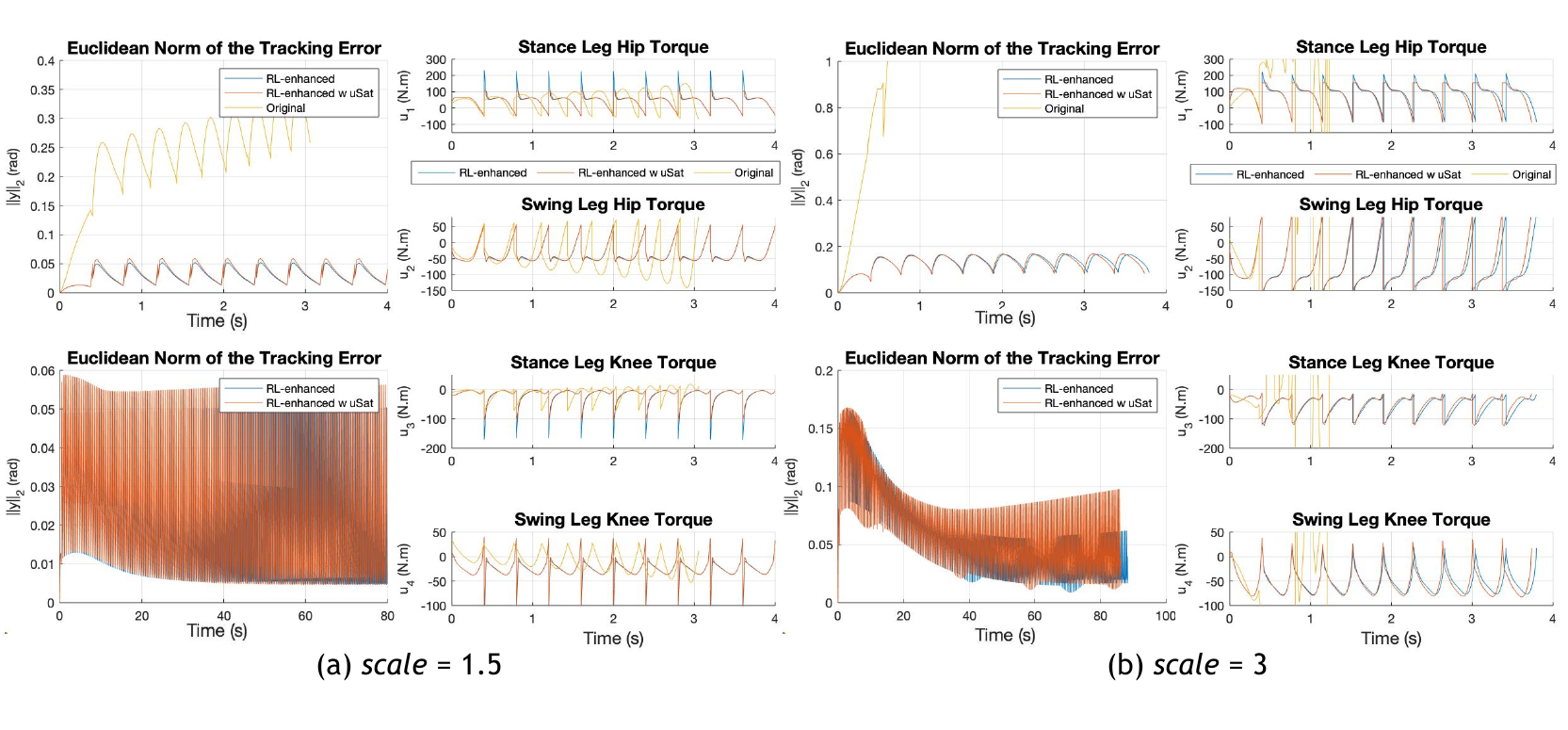}
    \vspace{-15pt}
    \caption{Euclidean norm of the tracking error (for 10 and 200 steps) and joint torques (for 10 steps), for the \textit{original IO controller} (yellow), the \textit{RL-enhanced IO controller} (blue), and the \textit{RL-enhanced IO controller} with torque saturation (red). Torque saturation for the \textit{RL-enhanced IO controller}  is set at $105$ $Nm$ when $scale=1.5$, and at $155$ $Nm$ when $scale=3$. There is no torque saturation for the \textit{original IO controller}.}
    \label{fig:scale153}
    \vspace{-2pt}
\end{figure}

\hspace{0.5cm} We introduce model uncertainty by scaling all the masses and inertia values of the plant's links by some factor (\emph{scale}) with respect to the known model. After about twenty minutes of training when the \emph{scale} is 1.5 and about an hour when the \emph{scale} is 3, we obtain the results shown in Figure \ref{fig:scale153}, in which we compare the tracking error and the joint torques when using (i) the \textit{original IO controller}, (ii) the \textit{RL-enhanced IO controller} without torque saturation and (iii) the \textit{RL-enhanced IO controller} when there is torque saturation. For these results we did not need to include torque saturation in the training process, and Figure \ref{fig:scale153} shows that the \textit{RL-enhanced IO controller} still performs well in the presence of input constraints if they are not too severe. The beneficial effects of including torque saturation constraints during training will be discussed later.

In Figure \ref{fig:scale153} it can be observed that the \textit{RL-enhanced IO controller} with and without saturation is able to stabilize the system indefinitely each time, whereas the \textit{original IO controller} accumulates error on the outputs and the robot falls after a few steps. Moreover, the \textit{RL-enhanced IO controller} achieves this without increasing the magnitude of the torques when compared with the \textit{original IO controller}.\smallbreak
The stability of the periodic gait obtained under the \textit{RL-enhanced IO controller} can also be studied by
the method of Poincar\'e. We consider the post-impact double stance surface $S$ as a
Poincar\'e section, and define the Poincar\'e map $P: S \rightarrow S$.
We can numerically calculate the eigenvalues of the linearization of the Poincar\'e map about the obtained periodic gait, which results in a dominant eigenvalue of magnitude $0.67$ for $scale = 1.5$ and no torque saturation, $0.78$ for $scale = 1.5$ with torque saturation, $0.76$ for $scale = 3$ and no torque saturation and $0.83$ for $scale = 3$ with torque saturation. The magnitude of the dominant eigenvalue being always less than one means that the designed controllers achieve local exponential stability \citep{westervelt2018feedback}.

\begin{figure}
    \centering
    \includegraphics[scale=0.38]{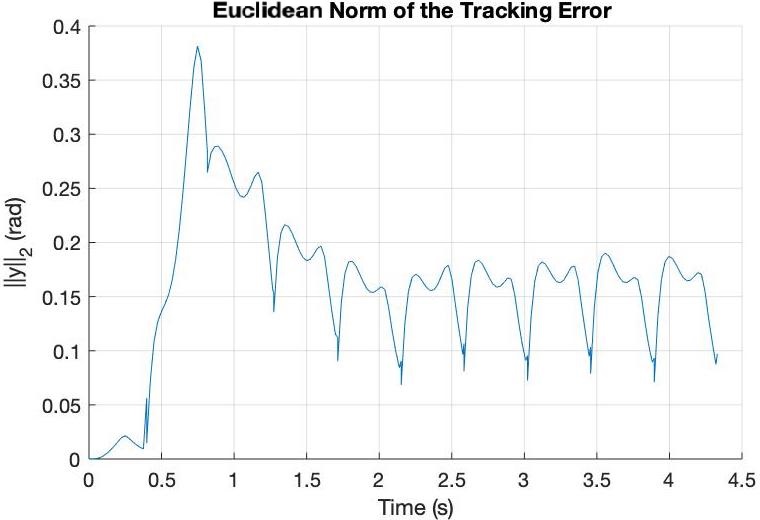}
    \includegraphics[scale=0.232]{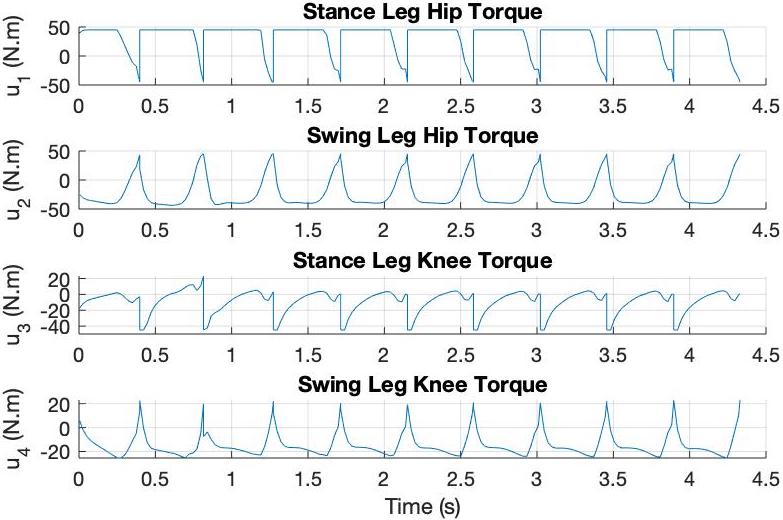}
    \vspace{-5pt}
    \caption{\textit{RL-enhanced IO controller} with torque saturation at $45$ $Nm$  and $scale=1$. Euclidean norm of the tracking error (left) and joint torques (right) for a simulation of 10 walking steps. The \textit{original IO controller} fails after one step and is not shown in this figure.}
    \vspace{-5pt}

\label{fig:mis1sat}

\end{figure}

\smallbreak
Next, we study the case of having no mismatch between the plant and the model dynamics but, instead, having heavy input constraints in the torques, which make the \textit{original IO controller} fail. By training while taking into account the torque saturation, we obtain a \textit{RL-enhanced IO controller} that achieves stable walking under the presence of severe input constraints, as shown in Figure \ref{fig:mis1sat}.

\subsection{Tracking Untrained Trajectories}
\begin{figure}
    \centering
    \includegraphics[scale=0.65]{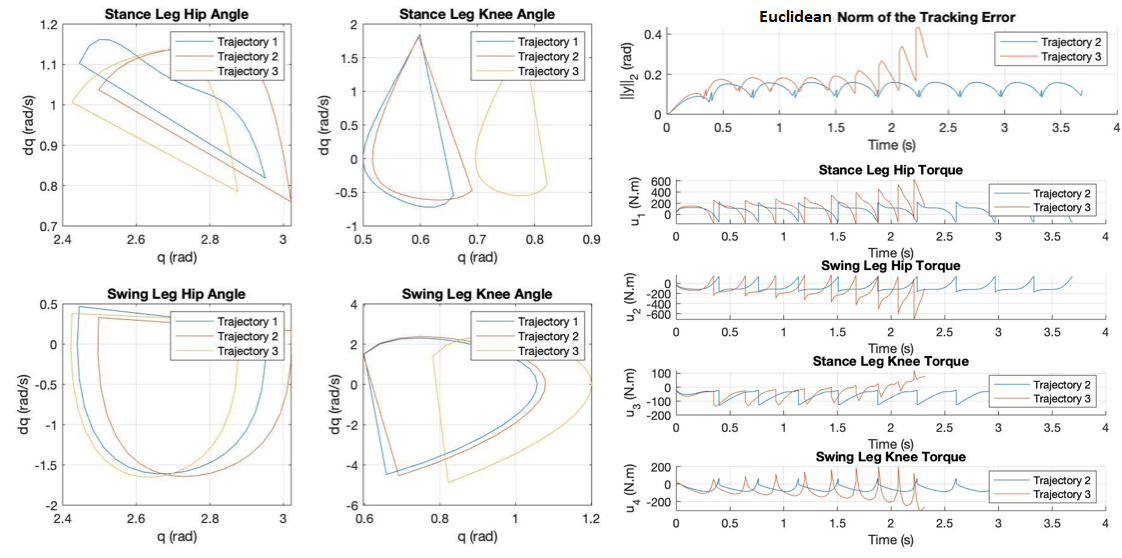}
    \vspace{-5pt}
    \caption{Left: Phase portrait of the periodic orbits. Right: Euclidean norm of the tracking error and joint torques for a simulation of 10 steps on untrained trajectories.}
    \label{fig:traj}
\end{figure}

\hspace{0.5cm} Depicted in Figure \ref{fig:traj} are the tracking errors and torques produced by the \textit{RL-enhanced IO controller} for a \emph{scale} of 3 when it is trying to follow periodic orbits it was not trained on. These trajectories differ from the one used for the training (\emph{trajectory 1}) in the maximum hip height during a step. As can be seen in the left part of Figure \ref{fig:traj}, \emph{trajectory 2} and \emph{trajectory 1} are relatively similar, whereas \emph{trajectory 3} constitutes a noticeably different walking gait. From the figures, we can see that the \textit{RL-enhanced IO controller} performs better when tested in \emph{trajectory 2} than in \emph{trajectory 3}. Actually, it will be able to stably track \emph{trajectory 2} for an indefinitely long horizon and not \emph{trajectory 3}. This was expected, since the more different the trajectory is, the farther the state of the robot will be from the distribution of states the DDPG agent has been trained on. Also, the output functions we have defined depend on the B\'ezier coefficients of the reference trajectory, and so the actual input-output linearizing controller is different for each trajectory. Still, thanks to training the DDPG agent on a stochastic distribution of initial states, we get enough exploration to achieve good tracking performance on untrained trajectories as long as they are not too different from the one the agent was trained on.

\section{Conclusions}
\label{Sec5}
\hspace{0.5cm} In this paper, we deployed a framework for improving an input-output linearizing controller for a bipedal robot when uncertainty in the dynamics and input constraints are present. We demonstrated the effectiveness of this approach by testing the learned controller on the hybrid, nonlinear and underactuated five-link walker RABBIT. For the simulations, different degrees of model-plant mismatch with and without torque saturation were used. Furthermore, the \textit{RL-enhanced IO controller} was able to follow trajectories it was not trained on as long as these trajectories were not too different from the one used for the training. However,
a limitation of our work is the need for the \textit{original IO controller} to work for a significant part of a walking step before failing, in order for the training process to converge. For high degrees of uncertainty this could be difficult to guarantee.
\smallbreak
Future work would focus on deploying this controller on hardware and on other more complex bipedal walkers, such as Cassie. Moreover, a similar approach could be used to improve Control Lyapunov Function (CLF)-based controllers in the presence of model uncertainty.
\vspace{-3pt}
\acks{The work of Fernando Casta\~neda was supported by a fellowship (code LCF/BQ/AA17/11610009) from ”la Caixa” Foundation (ID 100010434). This work was also partially supported through National Science Foundation Grants CMMI-1931853, IIS-1834557, by Berkeley Deep Drive and by HICON-LEARN (design of HIgh CONfidence LEARNing-enabled systems), Defense Advanced Research Projects Agency award number FA8750-18-C-010.}

\bibliography{ref}

\end{document}